\documentclass[aps,prd,twocolumn,preprintnumbers,nofootinbib,10pt]{revtex4-2}
\pdfoutput=1

\usepackage{comment}
\usepackage{graphicx}
\usepackage{slashed}
\usepackage[dvipsnames,table]{xcolor}
\usepackage[normalem]{ulem}
\usepackage{xspace}
\usepackage[tight]{subfigure}
\usepackage{amsmath}
\usepackage{amssymb}
\usepackage{amsfonts}
\usepackage{mathrsfs}
\usepackage{comment}
\usepackage{afterpage}
\usepackage{verbatim}
\usepackage{booktabs}
\usepackage{array}
\usepackage[title,titletoc]{appendix}
\usepackage{tabularx}
\usepackage{hyperref} 
\newcolumntype{C}[1]{>{\centering\arraybackslash}p{#1}}

\usepackage[dvipsnames]{xcolor}
    \definecolor{darkgreen}{rgb}{0,0.5,0}
    \definecolor{darkblue}{rgb}{0,0,0.6}
    \definecolor{purple}{rgb}{0.4,.2,0.7}
    \definecolor{medblue}{RGB}{0,23,127}
    \definecolor{teal}{RGB}{58,144,100}
\newcommand{\blue}[1]{\textcolor{medblue}{#1}}
\newcommand{\teal}[1]{\textcolor{teal}{#1}}

\def\citere#1{\mbox{Ref.~\cite{#1}}}

\newcommand{\newc}{\newcommand}
\newc{\beq}{\begin{equation}}
\newc{\eeq}{\end{equation}}
\newc{\beqn}{\begin{eqnarray}}
\newc{\eeqn}{\end{eqnarray}}
\newc{\bit}{\begin{itemize}}
\newc{\eit}{\end{itemize}}
\newc{\ben}{\begin{enumerate}}
\newc{\een}{\end{enumerate}}
\newc{\bce}{\begin{center}}
\newc{\ece}{\end{center}}
\newc{\bfi}{\begin{figure}}
\newc{\efi}{\end{figure}}

\newcommand{\rd}{\mathrm d}

\newcommand{\GeV}{\ensuremath{\,\text{GeV}}\xspace}
\newcommand{\TeV}{\ensuremath{\,\text{TeV}}\xspace}

\newcommand{\PH}{\ensuremath{{H}}\xspace}
\newcommand{\Pe}{\ensuremath{{e}}\xspace}
\newcommand{\PW}{\ensuremath{{W}}\xspace}
\newcommand{\PZ}{\ensuremath{{Z}}\xspace}

\newcommand{\Pj}{\ensuremath{\text{j}}\xspace}
\newcommand{\Pp}{\ensuremath{\text{p}}}

\newcommand{\Pt}{\ensuremath{\text{t}}\xspace}

\newcommand{\Mt}{\ensuremath{m_\Pt}\xspace}
\newcommand{\MH}{\ensuremath{M_\PH}\xspace}

\newcommand{\MW}{\ensuremath{M_\PW}\xspace}

\newcommand{\MZ}{\ensuremath{M_\PZ}\xspace}

\newcommand{\Gt}{\ensuremath{\Gamma_\Pt}\xspace}

\newcommand{\GZ}{\ensuremath{\Gamma_\PZ}\xspace}

\newcommand{\GW}{\ensuremath{\Gamma_\PW}\xspace}

\newcommand{\GF}{\ensuremath{G_\mu}}

\newcolumntype{.}{D{.}{.}{-1}}
\newcolumntype{d}[1]{D{.}{.}{#1}}
\colorlet{tableoverheadcolor}{gray!37.5}
\colorlet{tableheadcolor}{gray!25}
\colorlet{tablerowcolor}{gray!12.5}

\def\draftdate{\relax}
\def\mda{\relax}
\def\mua{\relax}
\def\mla{\relax}
\def\draft{
\def\thtystars{******************************}
\def\sixtystars{\thtystars\thtystars}
\typeout{}
\typeout{\sixtystars**}
\typeout{* Draft mode!
         For final version remove \protect\draft\space in source file *}
\typeout{\sixtystars**}
\typeout{}
\def\draftdate{\today}
\def\mua{\marginpar[\boldmath\hfil$\uparrow$]%
                   {\boldmath$\uparrow$\hfil}\color{black}%
                    \typeout{marginpar: $\uparrow$}\ignorespaces}
\def\mda{\color{red}\marginpar[\boldmath\hfil$\downarrow$]%
                   {\boldmath$\downarrow$\hfil}%
                    \typeout{marginpar: $\downarrow$}\ignorespaces}
\def\mla{\marginpar[\boldmath\hfil$\rightarrow$]%
                   {\boldmath$\leftarrow $\hfil}%
                    \typeout{marginpar: $\leftrightarrow$}\ignorespaces}
\def\Mua{\marginpar[\boldmath\hfil$\Uparrow$]%
                   {\boldmath$\Uparrow$\hfil}\color{black}%
                    \typeout{marginpar: $\uparrow$}\ignorespaces}
\def\Mda{\color{red}\marginpar[\boldmath\hfil$\Downarrow$]%
                   {\boldmath$\Downarrow$\hfil}%
                    \typeout{marginpar: $\downarrow$}\ignorespaces}
\def\Mla{\marginpar[\boldmath\hfil\textcolor{red}{$\Rightarrow$}]%
                   {\boldmath\textcolor{red}{$\Leftarrow $}\hfil}%
                    \typeout{marginpar: $\leftrightarrow$}\ignorespaces}
\overfullrule 5pt
\oddsidemargin 15mm
\marginparwidth 29mm
}

\newcommand{\mc}{\mathcal}

\begin{document}

\title{
Limitations of quantum tomography in Higgs-boson decays to four leptons} 
\author{Morgan Del Gratta$\,^{a,b}$,}\email{morgan.delgratta@unimi.it}
\author{Fabio Maltoni$\,^{c,d,e,f}$,}\email{fabio.maltoni@unibo.it}
\author{Davide Pagani$\,^{d}$}\email{davide.pagani@bo.infn.it}
\author{Giovanni Pelliccioli$\,^{g,h}$}\email{giovanni.pelliccioli@unimib.it}
\affiliation{\vspace*{0.3cm}$\,^{a}$University of Milano, Department of Physics, 20133 Milano, Italy}
\affiliation{$\,^{b}$INFN, Sezione di Milano, 20133 Milano, Italy}
\affiliation{$\,^{c}$University of Bologna, Department of Physics and Astronomy, 40126 Bologna, Italy}
\affiliation{$\,^{d}$INFN, Sezione di Bologna, 40126 Bologna, Italy}
\affiliation{$\,^{e}$European Organisation for Nuclear Research (CERN), 1211 Geneva, Switzerland}
\affiliation{$\,^{f}$Centre for Cosmology, Particle Physics and Phenomenology (CP3), Universit\'e Catholique de Louvain, B-1348 Louvain-la-Neuve, Belgium}
\affiliation{$\,^{g}$University of Milano--Bicocca, Department of Physics, 20126 Milano, Italy}
\affiliation{$\,^{h}$INFN, Sezione di Milano--Bicocca, 20126 Milano, Italy}


\begin{abstract}
We investigate the limitations of the two-qutrit quantum tomography
in Higgs-boson decays to four charged leptons.
At next-to-leading order, the standard reconstruction
can yield an operator that is not positive semidefinite,
even when radiatively improved spin-analysing powers are used.
We trace this failure to the coherent combination of photon-
and $\PZ$-mediated contributions to the off-shell dilepton system,
whose distinct production and decay structures invalidate
the assumed universal decay analysers.
Nevertheless, the rank-two angular sector remains accessible
and perturbatively stable, enabling a robust lower bound
on the squared concurrence, and hence a reliable assessment of entanglement
beyond leading order.
\end{abstract}

\keywords{LHC, quantum tomography, spin correlations, NLO EW, Higgs boson}
\preprint{COMETA-2026-37}

\vspace*{0.5cm}
\maketitle
\section{Introduction}\label{sec:intro}
The extraction of spin correlations and
quantum-information-inspired observables from LHC data
extends beyond top-antitop-quark pairs
\cite{ATLAS:2023fsd,CMS:2024pts,CMS:2024zkc},
which constitute two-qubit systems.
Recent ATLAS and CMS measurements in Higgs-boson decays
to four charged leptons also probe two-qutrit systems
\cite{CMS:2025anw,ATLAS:2026hye}. 
The theoretical understanding of suitable entanglement and Bell non-locality markers for two-qutrit spin systems at colliders 
has been developed extensively \cite{Barr:2021zcp,Aguilar-Saavedra:2022wam,Ashby-Pickering:2022umy,Aguilar-Saavedra:2022mpg,Fabbrichesi:2023cev,Fabbrichesi:2023jep,Aoude:2023hxv,Bernal:2023ruk,Fabbri:2023ncz,Morales:2023gow,Bernal:2024xhm,Grossi:2024jae,Sullivan:2024wzl,Wu:2024ovc,Grabarczyk:2024wnk,Aguilar-Saavedra:2024jkj,Subba:2024aut,DelGratta:2025qyp,Ding:2025mzj,Aguilar-Saavedra:2025byk,Goncalves:2025mvl,Ruzi:2025jql,Goncalves:2025xer,DelGratta:2025xjp,Aguilar-Saavedra:2025byk,Pelliccioli:2026ltl,Aguilar-Saavedra:2026wuq,Goncalves:2026njf,Chang:2026nzq},
although much of this work assumes a leading-order (LO) description of the underlying process. Tripartite entanglement in Higgs-boson decays has also been recently studied \cite{Morales:2024jhj,Aguilar-Saavedra:2024whi,Banacki:2026msu}.


In this work, we identify the origin and scope of the failure of
standard two-qutrit quantum tomography (QT) in Higgs-boson decays to
four charged leptons. Our starting point is the distinction between
extracting angular coefficients and interpreting them as entries
of a spin-density matrix: the latter requires a factorised
production--decay description with universal decay analysers.
Previous studies have found large next-to-leading-order (NLO) electroweak (EW) corrections
and negative eigenvalues in the reconstructed matrix for
$\PH\to \PZ\PZ^*\to4\ell$
\cite{Grossi:2024jae,DelGratta:2025qyp,Goncalves:2025mvl,Aguilar-Saavedra:2025byk}.
For processes dominated by two on-shell $Z$ bosons, much of
the effect can be absorbed into radiatively corrected
spin-analysing powers
\cite{Goncalves:2025mvl,DelGratta:2025xjp}.
We investigate why this prescription does not suffice when
one dilepton pair is far off shell.

Combining exact off-shell predictions with calculations in the
single-$\PZ$ narrow-width approximation, we test several possible
explanations. Higher angular harmonics remain negligible,
consistently with Ref.~\cite{Aguilar-Saavedra:2026wuq}, while
the reconstruction problem persists when resolved-photon effects
are suppressed and the off-shell dilepton virtuality is fixed.
Including squared one-loop amplitudes proves that the observed non-positivity is not merely an artefact
of perturbative truncation.
These checks show that an angular distribution restricted
to ranks $l\leq2$ does not by itself guarantee a consistent
two-qutrit reconstruction.

We identify the central obstruction as the coexistence of
$\PZ^*$- and $\gamma^*$-mediated contributions to the off-shell
dilepton current, with distinct production tensors and decay
analysers. Their coherent sum does not, in general, admit
the universal production--decay map assumed in standard QT.
We support this interpretation with two complementary tests.
First, replacing the off-shell charged-lepton pair
by neutrinos, which entails no $\gamma^*$-mediated contributions, restores a positive semidefinite and
perturbatively stable reconstruction.
The stability of this control channel also suggests
that loop contributions without an $s$-channel
$\PZ$ or photon coupling to the off-shell current
are numerically subleading in the observables
considered.
Second, an effective $HZ\gamma$ interaction in the Standard-Model Effective Field theory (SMEFT) reproduces
the failure already at tree level, also including terms
quadratic in the SMEFT coefficients.
The corresponding $HZZ$-only deformation preserves
the reconstruction.
The failure of the assumed tomographic map is there in the presence of both linear and quadratic EFT terms, and thus cannot be attributed solely to the SMEFT-series truncation, consistently with the perturbative expansion in the EW coupling within the SM.

Finally, we identify quantum information that remains
accessible despite this limitation.
The rank-two sector is independent of the rank-one
spin-analysing powers and receives only small
electroweak corrections.
Under the assumptions specified in
Sec.~\ref{sec:Cbound}, it allows us to construct
a perturbatively stable lower bound on entanglement
without reconstructing the problematic rank-one
spin correlations.
The corresponding rank-one angular moments remain
measurable, although their conversion into
spin-density coefficients is not universal.

The paper is organised as follows.
Section~\ref{sec:framework} introduces the angular
tomography formalism and its relation to two-qutrit
density-matrix reconstruction.
Section~\ref{sec:results} presents the numerical results
and analytical arguments that explain the limitations
of the standard reconstruction and identify robust
entanglement markers.
We draw our conclusions in Section~\ref{sec:conclusion}.

\section{The framework}\label{sec:framework}
We consider the decay of a SM Higgs boson into four charged leptons:
\begin{eqnarray}
\PH&\to&{\Pe}^+{\Pe}^-\,{\mu^+}{\mu^-}  \,.
\end{eqnarray}
We use the same notation and coordinate system as in \citere{DelGratta:2025xjp}, with decay angles $\{\theta_i,\phi_i\}_{i=1,2}$ associated to positively charged leptons and defined in the respective same-flavour lepton-pair rest frame, \emph{w.r.t.}~the lepton-pair flight direction in the overall Higgs-boson rest frame. 

We first characterise the angular distribution independently
of any spin-density-matrix interpretation. Retaining spherical
harmonics up to rank two in each dilepton system, we write:
\begin{align}\label{eq:VV}
&\left(\cfrac{\rd\sigma}{\rd\Phi}\right)^{\!\!-1}\!\!\!\frac{\rd \sigma}{
\rd\Phi\rd\Omega_{1}
\rd\Omega_{2}
}  = \\
&=\frac{1}{(4\pi)^2}
+ \frac{1}{4\pi}\, \sum_{l=1}^2 \sum_{m=-l}^{l}\, \alpha^{(1)}_{lm}(\Phi) \, Y_{lm}\big (\Omega_{1}\big ) \nonumber \\[2mm]
& \phantom{xx}+ \frac{1}{4\pi}\, \sum_{l=1}^2 \sum_{m=-l}^{l}\, \alpha^{(2)}_{lm}(\Phi) \, Y_{lm}\big (\Omega_{2}\big )
\nonumber \\[2mm]
& \phantom{xx}+ \sum_{l=1}^2 \sum_{l^\prime=1}^2 \sum_{m=-l}^{l}\sum_{m^\prime=-l^\prime}^{l^\prime}\, \gamma_{lm l^\prime m^\prime}(\Phi) \, Y_{lm}\big (\Omega_{1}\big ) \, Y_{l^\prime m^\prime}\big (\Omega_{2}\big ) \,,\nonumber
\end{align} 
where $\rd\Omega_i=\rd\!\cos\theta_{i}
\,\rd\phi_{i}$ for $i=1,2$, and $\Phi$ represents a generic dependence on production-level kinematics.
The truncation at $l,l'\leq2$ is exact for the tree-level
$\PH\to\PZ\PZ^*\to4\ell$ amplitude in the massless-lepton limit.
Beyond LO, its accuracy must be checked explicitly;
we examine higher-rank contributions in
Section~\ref{sec:highrank}.
The coefficients $\alpha$ and $\gamma$ are angular moments
and remain well-defined observables irrespective of whether
they admit a two-qutrit interpretation.

In the case of the Higgs-boson decay, $\rd\Phi \propto \rd Q_1^2\rd Q_2^2\rd\Theta$, where $Q_1,Q_2$ are the virtualities of the
lepton-pair systems, and $\Theta$ is the production angle of a single lepton pair in the Higgs-boson rest frame.

For full angular coverage, the coefficients can be
extracted by orthogonal projection
\cite{Aguilar-Saavedra:2022wam,Grossi:2024jae,DelGratta:2025qyp}.
For fiducial measurements, acceptance effects must
first be corrected for, or included explicitly in
the extraction procedure \cite{Grossi:2024jae}:
\begin{align}\label{eq:gam}
\alpha^{(i)}_{l m} &=
\displaystyle\frac1{\sigma}\int\!
\rd \Omega_i\, \cfrac{\rd \sigma}{
\rd \Omega_i}\,Y^*_{l m}(\Omega_i)  \,,\qquad i=1,2\nonumber\\
\gamma_{l ml'm'} &=
{
\displaystyle\frac1{\sigma}
\int\!
\rd \Omega_1\rd \Omega_2\, \cfrac{\rd \sigma}{
\rd \Omega_1\rd \Omega_2}\,Y^*_{l m}(\Omega_1)Y^*_{l' m'}(\Omega_2)\,.
}
\,
\end{align}
For simplicity, the remaining kinematic variables
$\Phi$ have been integrated out, but
the same projections can also be performed
differentially in $\Phi$.
A two-qutrit interpretation requires an additional ingredient:
the angular distribution must admit a representation in terms
of a positive semidefinite, unit-trace production density matrix
$\rho$ and decay matrices $\Gamma_{1,2}$ that are independent
of the production mechanism. With our normalisation,
this representation reads:
\begin{eqnarray}\label{eq:Rho}
&&
\frac{1}{\sigma}
\frac{\rd\sigma}{\rd\Omega_1\,\rd\Omega_2}
=
\frac{9}{16\pi^2}
\operatorname{Tr}\!\left[
\rho\,
\bigl(\Gamma_1(\Omega_1)\otimes\Gamma_2(\Omega_2)\bigr)^{\rm T}
\right]
\\
&&\qquad=\!\sum_{\lambda_1,\lambda_1',\lambda_2,\lambda_2'}\rho_{\lambda_1,\lambda_1',\lambda_2,\lambda_2'}
\Gamma_{\lambda_1,\lambda_1'}(\Omega_1)
\Gamma_{\lambda_2,\lambda_2'}(\Omega_2)\,,\nonumber
  \end{eqnarray}
where $\rho$ is the $9\times9$ spin-density matrix of interest and $\Gamma_{1(2)}$ is the $3\times3$ decay matrix associated with the first (second) qutrit ($\PZ$ boson, in our case).

When this representation holds, the angular moments can be
mapped onto the coefficients $A$ and $C$ of the spin-density
matrix through the decay analysing powers
\cite{Aguilar-Saavedra:2022wam,DelGratta:2025qyp}.
For non-vanishing $\eta_\ell^{(i)}$, the inverse map is:
\begin{align}\label{eq:gamma_to_C}
    A^{(i)}_{l m}=\xi^{(i)}_l \alpha^{(i)}_{l m}\,,\qquad C_{l m l'm'}=\xi^{(1)}_l\,\xi^{(2)}_{l'}\,\gamma_{l ml'm'}\,,\nonumber\\
    \xi^{(i)}_1 = \frac{\sqrt{8\pi}}{\eta_\ell^{(i)}}\,,\qquad \xi^{(i)}_2 = \sqrt{40\pi}\,,\qquad i=1,2    \,.
\end{align}

Equation~\eqref{eq:gamma_to_C} separates two logically distinct
steps: the extraction of angular moments and their conversion
into spin-density coefficients.
The rank-one conversion depends on the spin-analysing powers,
whereas the rank-two conversion involves only numerical factors.
At tree level, the $\PH\to\PZ\PZ^*\to4\ell$ amplitude provides
the factorised production--decay structure required for this
interpretation \cite{Aguilar-Saavedra:2022wam}.
Beyond LO, applying the same inverse map still defines
a Hermitian, unit-trace operator, but neither its positivity
nor the validity of the assumed decay analysers is guaranteed.
Our starting point is therefore to extract the angular moments
from the NLO-corrected distribution and test whether
they admit a consistent reconstruction, allowing also for
perturbatively improved spin-analysing powers.

\section{Results}\label{sec:results}
\subsection{Details of the calculations}\label{eq:details}
\begin{figure}[t]
    \centering
    \includegraphics[width=0.99\linewidth]{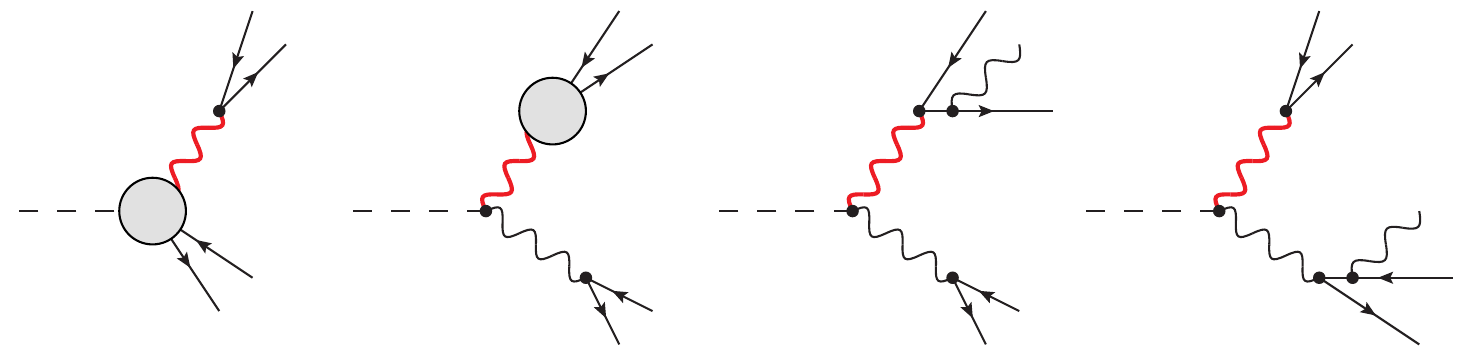}
    \caption{Factorisable one-loop and photon-emission diagram topologies 
    contributing to $\PH\to\PZ(\Pe^+\Pe^-)\mu^+\mu^-$ at NLO EW in the 
    narrow-width approximation, with $\MH<2\MZ$. The intermediate, on-shell $\PZ$ boson is highlighted in red color, while the off-shell one (when present) is in black colour. The gray blobs stand for renormalised one-loop contributions, while the black dots for tree-level couplings.}
    \label{fig:diagsH125}
\end{figure}

We use two complementary calculations. Full off-shell
predictions for $\PH\to\Pe^+\Pe^-\mu^+\mu^-$ at NLO EW
accuracy are obtained with {\sc MadGraph5\_aMC@NLO}
\cite{Alwall:2014hca,Frederix:2018nkq}, including all
resonant and non-resonant contributions and both real
and virtual corrections.
A standalone Monte Carlo code interfaced with {\sc Recola}
\cite{Actis:2016mpe} provides predictions in the strict
single-$\PZ$ narrow-width approximation (NWA),
$\PH\to\PZ(\Pe^+\Pe^-)\mu^+\mu^-$.
Further validation is performed in the pole approximation
with {\sc MoCaNLO} \cite{Denner:2026phn}.
The relevant diagram topologies are shown in
Fig.~\ref{fig:diagsH125}.

The full calculation establishes the behaviour of the
physical angular observables, including photon radiation
and lepton dressing. The single-NWA calculation isolates
the on-shell $\PZ$ decay and allows us to investigate
the off-shell dilepton current at fixed virtuality,
using Born-like kinematics.

In the single-NWA calculation, we construct virtual
approximations from the UV-renormalised one-loop amplitude
after minimal infrared subtraction
\cite{Catani:1998bh,Schonherr:2017qcj}.
The NLO$_{\rm virt}$ prediction includes the Born
contribution and its interference with the finite one-loop
amplitude. The aNNLO$_{\rm virt}$ prediction additionally
includes the square of that one-loop amplitude.
Both predictions are evaluated with Born-like kinematics
and contain no resolved real photons.
We stress that the aNNLO$_{\rm virt}$ result is not a complete NNLO
prediction: genuine two-loop contributions and the
corresponding real-radiation terms are not included.
Its purpose is to test the role of squared one-loop
contributions in the angular moments and in the positivity
of the reconstructed operator.

The following numerical input parameters are used.
A SM Higgs boson with $\MH = 125\GeV$ is considered
throughout the whole paper.
We use following pole masses and widths for EW bosons:
\footnote{Note that in the single-NWA calculations 
(for an intermediate on-shell $\PZ$ boson) 
we also set $\GW=0$, 
so to satisfy EW gauge invariance.}
\begin{eqnarray}
  \MW &=   80.34997\GeV,\quad
  \GW &= 2.0843\GeV, \quad\nonumber\\
  \MZ &=   91.15348\GeV,\quad 
  \GZ &= 2.4946\GeV\,,\nonumber
\end{eqnarray}
the top-quark mass and width are set to
  \begin{equation}
  \Mt =   173\GeV,\quad
  \Gt =   1.36\GeV\,,
  \nonumber  
  \end{equation}
and the value of the EW coupling is obtained in the $G_\mu$ scheme \cite{Sirlin:1980nh,Denner:2000bj}, 
employing the following value for the Fermi constant:
\begin{equation}
  \GF = 1.16638\cdot 10^{-5} \GeV^{-2}.\nonumber
\end{equation}

Unless otherwise stated, we label the two dilepton systems
by flavour: system~1 is the $\Pe^+\Pe^-$ pair and system~2
the $\mu^+\mu^-$ pair. This assignment does not presume
a unique intermediate boson for either pair.
We select the electron pair near the $\PZ$ resonance
and impose a lower invariant-mass cut on the muon pair
to exclude the photon-pole region in the massless-lepton
calculation:
\begin{equation}\label{eq:select}
81\GeV<M_{\Pe^+\Pe^-}<101\GeV,
\quad Q=M_{\mu^+\mu^-}>5\GeV.
\end{equation}
In the single-NWA calculation,
$M_{\Pe^+\Pe^-}=\MZ$ and $Q\leq\MH-\MZ$.
In the full off-shell calculation, the corresponding
kinematic bound is $Q\leq\MH-M_{\Pe^+\Pe^-}$.
When real photons are included, the selections are
applied to dressed leptons.



\subsection{Higher-rank contributions}\label{sec:highrank}

We first test whether angular structures beyond the
$l,l'\leq2$ expansion in Eq.~\eqref{eq:VV} could account
for the failure of the standard reconstruction.
Such contributions would signal that the assumed
two-body spin-one decay description does not capture
the full angular dependence.

In the full off-shell calculation at NLO EW accuracy,
the higher-rank coefficients examined are negligible,
with only tiny non-vanishing contributions to the
single-system $l=4$, $m=0$ moments.
This agrees with the findings of
Ref.~\cite{Aguilar-Saavedra:2026wuq}, where higher-rank
coefficients were compatible with zero within numerical
uncertainties.
In the single-NWA calculation, we find no statistically
significant higher-rank coefficients at either
NLO$_{\rm virt}$ or aNNLO$_{\rm virt}$ accuracy.
The latter provides an additional check, since it
includes squared one-loop contributions with
intermediate-state structures absent at Born level.

These results support the numerical adequacy of the
rank-two angular expansion for the predictions considered.
They do not, however, establish a two-qutrit
density-matrix interpretation: restricting the angular
ranks does not guarantee that the coefficients can be
represented by a positive semidefinite production
density matrix and universal decay analysers.
As shown below, the reconstruction problem persists
even where higher-rank contributions are consistent
with zero.

\subsection{Integrated results}
The scalar nature of the Higgs boson strongly
constrains the angular distribution
\cite{Grossi:2024jae,DelGratta:2025qyp,Goncalves:2025mvl}.
After integration over the phase space selected
by Eq.~\eqref{eq:select}, we find seven independent
non-vanishing coefficients within the
$l,l'\leq2$ expansion, accounting for the symmetry
relation
\begin{eqnarray}\label{eq:symm}
    \gamma_{l ml-m}&=&\gamma_{l -ml m}\,, \qquad l=1,2\,.
\end{eqnarray}
which holds at both LO and NLO EW accuracy
\cite{Grossi:2024jae,DelGratta:2025qyp,Goncalves:2025mvl}.
The exact NLO EW results are shown in 
Tab.~\ref{tab:qtom_fulloffshell_125_GP} for two different lepton-dressing resolution radii.
\begin{table}[h!]
\begin{center}
\begin{tabular}{ crrr}
\hline
\multicolumn{4}{c}{$\PH\to \Pe^+\Pe^-\mu^+\mu^-$, integrated over $Q$}\\
\hline
  & LO & NLO (${\Delta R=0.1}$) & NLO ($\Delta R=1$) \\ \hline 
{$\alpha^{(1)}_{20}$} & $-$0.0482(1)  &  $-$0.0478(1) &$-$0.0484(2)\\
{$\alpha^{(2)}_{20}$} & $-$0.0481(1)   &    $-$0.0454(1) & $-$0.0480(2)\\
{$\gamma_{1010}$} &  $-$0.00111(1)  &  0.00029(1)&  0.00029(3)\\
{$\gamma_{111-1}$} & 0.00179(1) & $-$0.00031(1)& $-$0.00033(3)\\
{$\gamma_{222-2}$} & 0.00490(3) & 0.00485(4)& 0.00477(4) \\
 {$\gamma_{212-1}$} & $-$0.00773(2) & $-$0.00772(4)& $-$0.00773(6)\\
{$\gamma_{2020}$}  & 0.01098(4)  &  0.01090(4) & 0.01099(4)\\
\hline
\end{tabular}
\caption{
Results of the complete QT of $\PH\to \Pe^+\Pe^-\mu^+\mu^-$ for $\MH=125\GeV$ at LO and at exact NLO EW. At NLO two values of the photon-recombination resolution have been used ($\Delta R = 0.1,\,1$). 
The cuts $5\GeV<Q=M_{\mu^+\mu^-}<\MH-\MZ$, $81\GeV<M_{\Pe^+\Pe^-}<101\GeV$ are applied to dressed leptons.}
\label{tab:qtom_fulloffshell_125_GP}
\end{center}
\end{table}
Looking at the single-boson coefficients, we observe that the
LO symmetry relation,
\beq\label{eq:symm2}
\alpha_{20}^{(1)} =
\alpha_{20}^{(2)}\,,
\eeq
is approximately recovered for the larger dressing
radius considered.
For a smaller radius, unrecombined photons modify
the dilepton angular distributions and produce
a visible difference between the two coefficients.
This difference can also be reduced by vetoing
resolved photons with energies above
$E_\gamma^{\rm max}$ \cite{Aguilar-Saavedra:2026wuq},
as illustrated in Tab.~\ref{tab:qtom_alphas20_125_dr01}.
\begin{table}[h!]
\begin{center}
\begin{tabular}{ crrrr}
\hline
 $E^{\rm max}_{\gamma}  $                    & no cut & $10\GeV$ &$1\GeV$  & $0.1\GeV$ \\ \hline 
{$\alpha^{(1)}_{20}$} & $-0.0478(2)$ & $-0.0485(3)$ & $-0.0489(4)$ & $-0.0495(8)$ \\
{$\alpha^{(2)}_{20}$} & $-0.0457(2)$ & $-0.0461(3)$ & $-0.0480(4)$ & $-0.0488(7)$ \\
\hline
\end{tabular}
\caption{Single-qutrit coefficients in $\PH\to \Pe^+\Pe^-\mu^+\mu^-$ for $\MH=125\GeV$ at exact NLO EW, for a dressing radius $\Delta R = 0.1$. The cuts $M_{\mu^+\mu^-}>5\GeV$, $81\GeV<M_{\Pe^+\Pe^-}<101\GeV$ (applied to dressed leptons), and a further cut on the maximum unclustered-photon energy $E_{\gamma}<E^{\rm max}_{\gamma}$ are understood.}
\label{tab:qtom_alphas20_125_dr01}
\end{center}
\end{table}
The correlation moments show a marked contrast
between the two angular sectors.
The $\gamma_{2m2m'}$ coefficients receive small
NLO EW corrections, whereas the
$\gamma_{1m1m'}$ coefficients undergo large changes,
including sign reversals
\cite{Grossi:2024jae,DelGratta:2025qyp,Goncalves:2025mvl}.
For processes dominated by two on-shell $\PZ$ bosons,
much of this effect can be absorbed into corrected
spin-analysing powers
\cite{Goncalves:2025mvl,DelGratta:2025xjp}.
In the Higgs decay considered here, this prescription
is well established for the nearly on-shell decay.
Whether an analogous prescription exists for the
off-shell lepton pair requires a separate test.

To assess the role of resolved radiation and
of the finite virtuality spread of the resonant
electron pair, we compare these results with
the single-NWA virtual approximations described
in Sec.~\ref{eq:details}.
The corresponding angular coefficients are shown
in Tab.~\ref{tab:qtom_nwa_125}.
\begin{table}[h!]
\begin{center}
\hspace{-0.3cm}\begin{tabular}{ crrr}
\hline
\multicolumn{4}{c}{$\PH\to \PZ(\Pe^+\Pe^-)\mu^+\mu^-$, integrated over $Q$}\\
\hline
               & LO & NLO$_{\rm virt}$ & aNNLO$_{\rm virt}$ \\ 
\hline %
{$\alpha^{(1)}_{20}$}  & $-0.04788 (4 )$ & $-0.04802 (6 )$ &$-0.04732 (7 )$\\
{$\alpha^{(2)}_{20}$}  & $-0.04788 (4 )$ & $-0.04816 (6 )$ & $-0.04745 (6 )$\\
{$\gamma_{1010}$}  & $-0.00112 (1 )$ & $0.00035 (1 )$ & $0.00017 (2 )$\\
{$\gamma_{111-1}$}  & $0.00176 (2 )$ & $-0.00042 (1 )$ & $-0.00016 (2 )$\\
{$\gamma_{222-2}$}  & $0.00493 (2 )$ & $0.00493 (2 )$ & $0.00497 (2 )$\\
{$\gamma_{212-1}$} & $-0.00773 (1 )$ & $-0.00770 (1 )$ &$-0.00769 (2 )$\\
{$\gamma_{2020}$}  & $0.01098 (1 )$ & $0.01099 (1 )$ &$0.01095 (2 )$\\
\hline
\end{tabular}
\caption{
Results of the complete QT of $\PH\to \PZ(\Pe^+\Pe^-)\mu^+\mu^-$ for $\MH=125\GeV$
at LO and at approximate NLO and NNLO EW accuracy (NLO$_{\rm virt}$, aNNLO$_{\rm virt}$) in the single NWA, for the whole available range of $M_{\mu^+\mu^-}=Q$, $5\GeV<Q<\MH-\MZ$.
}\label{tab:qtom_nwa_125}
\end{center}
\end{table}

The NWA results at NLO$_{\rm virt}$ accuracy reproduce pretty well the coefficients obtained at exact NLO accuracy with $\Delta R = 1$, with percent or subpercent deviations for all $l=2$ coefficients.  Slightly larger deviations (some percent) is found for $l=1$ coefficients, but still compatible within numerical uncertainties. Such coefficients are very close to zero at NLO, leading to rather large numerical instabilities in their evaluation.
The inclusion of approximate NNLO$_{\rm virt}$ effects has a negligible impact on $l=2$ coefficients, while large effects are found for $l=1$ ones which get even closer to zero at this accuracy.

Overall, the NWA results suggest that the dramatically large corrections
to $l=1$ angular coefficients come from virtual effects associated to the off-shell lepton pair.

\subsection{Dependence on the lepton-pair virtuality}
The conversion of the angular coefficients
$\gamma_{1m1m'}$ into their spin-density counterparts
requires spin-analysing powers for both lepton pairs.
For the on-shell $\PZ$ decay, $\eta_{\ell}(\MZ)$ is
well defined and can be improved perturbatively
\cite{DelGratta:2025xjp}.
In the SM we obtain:
\begin{eqnarray}
    \eta_{\ell}^{\rm LO}(\MZ) &=&0.2131\,,\nonumber\\
    \eta_{\ell}^{\rm NLOEW}(\MZ) &=&0.1409\,,\nonumber\\
    \eta_{\ell}^{\rm aNNLOEW_{\rm virt}}(\MZ) &=&0.1401\,.
\end{eqnarray}
The numerical uncertainty affects the last decimal place.
The NLO EW correction reduces the analysing power by
approximately $34\%$ relative to LO, while the inclusion
of squared one-loop contributions produces only a small
additional shift.
This comparison tests the impact of the retained
higher-order terms, but does not establish the size
of the complete NNLO EW correction, which also requires
genuine two-loop contributions.

For the off-shell lepton pair, the situation is less
straightforward.
At LO, the $\PZ$-mediated decay current has the same
spin-analysing power as the on-shell decay, independently
of $Q$ in the massless-lepton limit.
Beyond LO, one may attempt to introduce an effective
analysing power $\eta_{\ell,\rm eff}$.
However, neither its independence of $Q$ nor its
ability to describe all rank-one angular coefficients
with a single value is guaranteed.

Studying the angular distribution at fixed $Q$ separates
this issue from the effects of integrating over
different dilepton virtualities.
In particular, such integration can mix different
LO spin states \cite{DelGratta:2025qyp}.
Fixing $Q$ removes this source of mixing, but does
not by itself guarantee purity or the validity of
the tomographic reconstruction beyond LO.

\begin{figure}[t]
    \centering
    \includegraphics[width=0.98\linewidth]{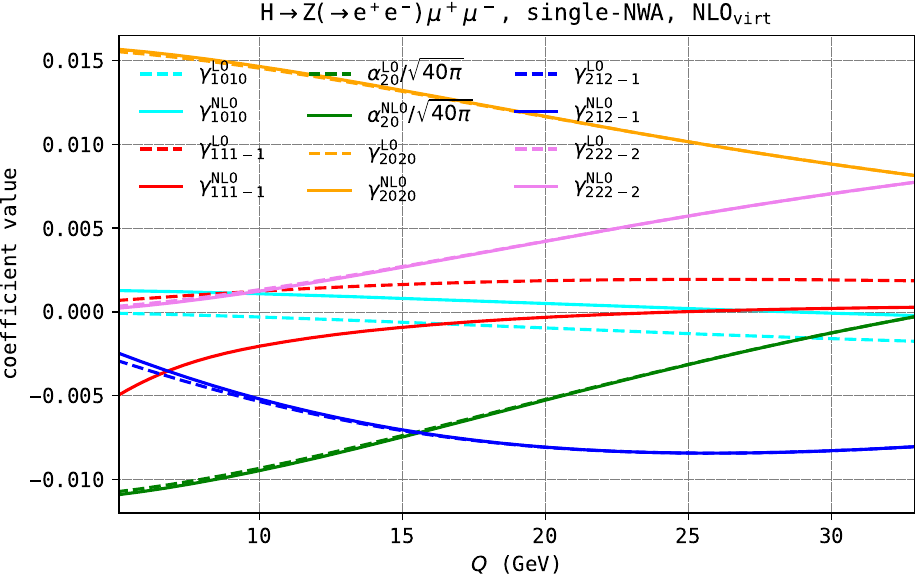}
    \caption{Dependence of the non-vanishing angular
    coefficients $\alpha_{\ell m},\,\gamma_{\ell m\ell' m'}$,
    $\ell=1,2$, on the virtuality $Q$ of the off-shell
    lepton pair ($\mu^+\mu^-$), for $\MH=125\GeV$,
    at LO and approximate NLO EW accuracy
    (NLO$_{\rm virt}$) in the single NWA.}
    \label{fig:q2dep}
\end{figure}

We therefore scan the non-vanishing angular coefficients
as functions of $Q$ at LO and NLO$_{\rm virt}$ accuracy
in the single NWA.
The results are shown in Fig.~\ref{fig:q2dep}.
We have also checked that the higher-rank contributions
discussed in Sec.~\ref{sec:highrank} remain compatible
with zero within numerical accuracy at the individual
values of $Q$ considered.

The rank-two coefficients receive small or negligible
NLO corrections throughout the range considered.
By contrast, both $\gamma_{1010}$ and $\gamma_{111-1}$
show substantial changes.
At LO, $\gamma_{1010}$ approaches zero at small
virtualities and becomes increasingly negative as
$Q$ grows.
The NLO$_{\rm virt}$ result follows a similar trend,
with an approximately positive shift of $0.0015$,
and crosses zero around $Q=25\GeV$.
The coefficient $\gamma_{111-1}$ is positive and
varies only mildly at LO.
At NLO$_{\rm virt}$, it instead takes sizeable negative
values at low $Q$, crosses zero around $25\GeV$,
and remains small and positive at larger virtualities.

These results show that the large corrections to
the rank-one sector persist locally in $Q$ and
are not generated solely by integration over the
dilepton virtuality.
The contrast with the perturbatively stable rank-two
sector motivates a direct consistency test:
at a fixed value of $Q$, can a single effective
analysing power $\eta_{\ell,\rm eff}(Q)$ recover
a physical density matrix?
We address this question in the following subsection.

\subsection{Testing the reconstruction at fixed virtuality}\label{sec:SMNLO}

We now test whether a consistent reconstruction can
be recovered at fixed virtuality by introducing
an effective analysing power for the off-shell pair.
We first examine the relations among angular
coefficients that do not depend on this choice.
The symmetry relation in Eq.~\eqref{eq:symm}
continues to hold at fixed $Q$ within our numerical
accuracy. We also find that the single-system
rank-two coefficients satisfy
\begin{eqnarray}
A^{(1)}_{20}(Q)=A^{(2)}_{20}(Q)=A_{20}(Q)\,.
\label{eq:relA}
\end{eqnarray}
Other relations amongst $l=2$ coefficients that are preserved by NLO corrections are (already written in terms of $A,\,C$ coefficients):
\begin{eqnarray}
C_{2020}(Q)&\displaystyle=1-\frac{A_{20}(Q)}{\sqrt{2}}\,,\nonumber\\
C_{222-2}(Q)&=\displaystyle1+\frac{A_{20}(Q)}{\sqrt{2}}\,.
\label{eq:relC}
\end{eqnarray}
These relations test the persistence of the LO
rank-two angular structure.
They do not establish purity or guarantee that
the complete distribution admits a universal
production--decay factorisation.
To test the rank-one sector, we fix the on-shell
analysing power to $\eta_\ell(\MZ)$ and ask whether
a single $\eta_{\ell,\rm eff}(Q)$ can simultaneously
satisfy the two LO consistency relations
\cite{DelGratta:2025qyp},
\begin{eqnarray}\label{eq:symmCrit}
    C_{1010}(Q)&=&C_{2020}(Q)-2\,,\nonumber\\
    C_{111-1}(Q)&=&-C_{212-1}(Q)\,.\\ \nonumber
\end{eqnarray}
At NLO neither of these two relations are guaranteed to be valid.
Therefore we parametrise the deviation from Eq.~\eqref{eq:symmCrit} as,
\begin{eqnarray}
\delta_0(Q)&\equiv& C_{1010}(Q)-\left[C_{2020}(Q)-2\right]\label{delta0}\,,\\
\delta_1(Q)&\equiv& C_{111-1}(Q)+C_{212-1}(Q)\,,
\end{eqnarray}
where we have made explicit the dependence on the specific value of $Q$ for the lepton-pair virtuality.

As a first reconstruction prescription, we impose
$\delta_1(Q)=0$ to define the effective analysing
power. This relation then holds by construction;
the independent diagnostic is whether the same
choice also gives $\delta_0(Q)=0$.
We obtain
\begin{eqnarray}
C_{111-1}(Q)&\equiv& -40\pi\gamma_{212-1}(Q)\,,\nonumber\\
C_{1010}(Q)&\equiv&
\frac{8\pi\,\gamma_{1010}(Q)}{\eta_{\rm eff}(Q)\,\eta_{\ell}(\MZ)}\,,
\end{eqnarray}
where the effective spin analysing power is defined as:
\begin{eqnarray}
{\eta_{\rm eff}(Q) \equiv-
\frac{\gamma_{111-1}(Q)} { 5\,\gamma_{212-1}(Q)\,\eta_{\ell}(\MZ)} }\,.
\end{eqnarray}
In this scheme, the deviation $\delta_0(Q)$ in Eq.~\eqref{delta0} becomes:
\begin{eqnarray}
\delta_0(Q)\equiv 2 - 40 \pi \left( \gamma_{2020}(Q) +\frac{\gamma_{1010}(Q) 
 }{\gamma_{111-1}(Q)} \gamma_{212-1}(Q) \right)\,.\nonumber\\[-0.2cm]
\end{eqnarray}
In this prescription, $\delta_0(Q)$ is expressed
entirely in terms of angular moments and is
independent of the on-shell analysing power.
A nonzero result therefore cannot be removed
by improving $\eta_\ell(\MZ)$ alone.
The prescription requires non-vanishing moments
in the denominators and becomes ill-defined
near their zero crossings.

We choose the virtuality $Q$ to be 10$\GeV$. The non-vanishing angular coefficients are shown in the upper block of Tab.~\ref{tab:qtom_nwa_125_vvll}. Both $l=1$ coefficients feature a change of sign going from LO to NLO, implying that an effective spin analysing power at this value has to change sign as well, in order to have perturbatively stable $C$ coefficients. 
Note that, although less populated compared to the region $Q\approx 30\GeV$ (in terms of total decay rate), decay events with lower virtualities (like $5\GeV\lesssim Q\lesssim 20\GeV$) give a sizeable, non-vanishing contribution to angular coefficients with $l=1$ which are instead zero in the most populated region (see Fig.~\ref{fig:q2dep}).

\begin{table}[h!]
\begin{center}
\hspace{-0.3cm}
\begin{tabular}{ crrr}
\hline
\multicolumn{4}{c}{$\PH\to \PZ(\Pe^+\Pe^-)\,\mu^+\mu^-$}\\
\hline
   $Q=10\GeV$                    & LO & NLO$_{\rm virt}$ & aNNLO$_{\rm virt}$\\ 
\hline
{$\alpha^{(1)}_{20}$}  & $-0.10473 (5 )$ & $-0.10611 (5 )$ & $-0.10243 (5 )$ \\
{$\alpha^{(2)}_{20}$}  & $-0.10474 (5 )$ & $-0.10631 (5 )$ & $-0.10260 (5 )$ \\
{$\gamma_{1010}$}  & $-0.00031 (1 )$ & $0.00108 (1 )$ & $0.00086 (1 )$ \\
{$\gamma_{111-1}$}  & $0.00122 (3 )$ & $-0.00207 (3 )$ & $-0.00157 (3 )$ \\
{$\gamma_{222-2}$}  & $0.00135 (3 )$ & $0.00126 (3 )$ & $0.00149 (3 )$ \\
{$\gamma_{212-1}$}   & $-0.00535 (1 )$ & $-0.00520 (1 )$ & $-0.00521 (1 )$ \\
{$\gamma_{2020}$}  & $0.01457 (1 )$ & $0.01467 (1 )$ & $0.01444 (1 )$ \\
\hline
\multicolumn{4}{c}{$\PH\to \PZ(\bar{\nu}_{\Pe}\nu_{\Pe})\,\mu^+\mu^-$}\\
\hline
   $Q=10\GeV$                    & LO & NLO$_{\rm virt}$  & aNNLO$_{\rm virt}$ \\ 
\hline
{$\alpha^{(1)}_{20}$}  & $-0.10472 (24 )$ & $-0.10608 (23 )$ & $-0.10243 (23 )$ \\
{$\alpha^{(2)}_{20}$}  & $-0.10472 (23 )$ & $-0.10628 (23 )$ & $-0.10260 (23 )$ \\
{$\gamma_{1010}$} & $-0.00145 (5 )$ & $0.00450 (5 )$ & $0.00451 (5 )$ \\
{$\gamma_{111-1}$}  & $0.00576 (13 )$ & $-0.00761 (13 )$ & $-0.00765 (13 )$ \\
{$\gamma_{222-2}$}  & $0.00132 (11 )$ & $0.00123 (11 )$ & $0.00146 (11 )$ \\
{$\gamma_{212-1}$}  & $-0.00534 (7 )$ & $-0.00520 (7 )$ & $-0.00521 (7 )$ \\
{$\gamma_{2020}$} & $0.01455 (6 )$ & $0.01465 (6 )$ & $0.01442 (6 )$  \\
\hline
\multicolumn{4}{c}{$\PH\to \PZ(\Pe^+\Pe^-)\,\nu_\mu \bar{\nu}_\mu$}\\
\hline
   $Q=10\GeV$                    & LO & NLO$_{\rm virt}$& aNNLO$_{\rm virt}$ \\ 
\hline
{$\alpha^{(1)}_{20}$} & $-0.10473 (9 )$ & $-0.10452 (9 )$ & $-0.10452 (9 )$ \\
{$\alpha^{(2)}_{20}$}  & $-0.10473 (10 )$ & $-0.10463 (10 )$ & $-0.10462 (10 )$ \\
{$\gamma_{1010}$}  & $-0.00145 (2 )$ & $-0.00098 (2 )$ & $-0.00097 (2 )$ \\
{$\gamma_{111-1}$}  & $0.00572 (4 )$ & $0.00383 (4 )$ & $0.00380 (4 )$ \\
{$\gamma_{222-2}$}  & $0.00135 (6 )$ & $0.00136 (6 )$ & $0.00136 (6 )$ \\
{$\gamma_{212-1}$}  & $-0.00534 (3 )$ & $-0.00536 (3 )$ & $-0.00536 (3 )$ \\
{$\gamma_{2020}$}  & $0.01456 (2 )$ & $0.01456 (2 )$ & $0.01456 (2 )$ \\
\hline
\end{tabular}
\caption{
Results of the complete QT of $\PH\to \PZ(\Pe^+\Pe^-)\mu^+\mu^-$ (top block), $\PH\to \PZ(\bar{\nu}_{\Pe}\nu_{\Pe})\mu^+\mu^-$ (middle block)
and $\PH\to \PZ(\Pe^+\Pe^-)\nu_\mu \bar{\nu}_\mu$ (bottom block)
for $\MH=125\GeV$
at LO and at approximate NLO and NNLO EW accuracy in the single NWA, for the specific value $Q=10\GeV$ of the off-shell fermion pair. The angular measure of the neutrino--antineutrino pair is associated to the antineutrino kinematics, for consistency with the leptonic case where angles are those of positively charged leptons (antiparticles).}\label{tab:qtom_nwa_125_vvll}
\end{center}
\end{table}

At approximate NLO in the NWA (at fixed $Q=10\GeV$), the reconstructed $\rho$ matrix reads (numerical uncertainties are on the fourth digit after the comma):
\begin{equation*}
\label{NLOrho10}
\begin{pmatrix}
\teal{~-0.030} & \cdot& \cdot& \cdot& \cdot& \cdot& \cdot& \cdot& \cdot\\
 \cdot& \cdot & \cdot& \cdot & \cdot& \cdot& \cdot& \cdot& \cdot\\
 \cdot& \cdot& \blue{0.083} & \cdot& \blue{-0.218} & \cdot& \blue{0.053} & \cdot& \cdot\\
 \cdot& \cdot & \cdot& \cdot & \cdot& \cdot& \cdot& \cdot& \cdot\\
 \cdot& \cdot& \blue{-0.218} & \cdot& \blue{0.895} & \cdot& \blue{-0.218} & \cdot& \cdot\\
 \cdot& \cdot& \cdot& \cdot& \cdot& \cdot & \cdot& \cdot & \cdot\\
 \cdot& \cdot& \blue{0.053} & \cdot& \blue{-0.218} & \cdot& \blue{0.057} & \cdot& \cdot\\
 \cdot& \cdot& \cdot& \cdot& \cdot& \cdot & \cdot& \cdot & \cdot\\
 \cdot& \cdot& \cdot& \cdot& \cdot& \cdot& \cdot& \cdot& \teal{-0.030~} \\
\end{pmatrix}\,. \nonumber
\end{equation*}
The negative diagonal entries demonstrate that
the reconstructed operator is not positive
semidefinite.
Imposing $\delta_1(Q)=0$ leaves a nonzero,
negative $\delta_0(Q)$, which generates
the unphysical diagonal populations.

The alternative prescription, $\delta_0(Q)=0$,
does not restore positivity either.
It leaves $\delta_1(Q)\neq0$, producing non-zero
entries $\rho_{24}$ and $\rho_{68}$ and their
Hermitian conjugates, therefore leading to negative eigenvalues for $\rho$.

\medskip

To determine which dilepton system drives the
failure, we perform two complementary control
tests, replacing either the on-shell or the
off-shell charged-lepton pair by neutrinos:
\begin{eqnarray}
\PH&\to&\PZ(\nu_{\Pe}\bar{\nu}_{\Pe})\,{\mu^+}{\mu^-}  \,,\\ \PH&\to&\PZ({\Pe^+}{\Pe^-})\,\nu_{\mu}\bar{\nu}_{\mu} \,.
\end{eqnarray}
The non-vanishing coefficients for these two final-state signatures are reported in the middle and lower blocks of Tab.~\ref{tab:qtom_nwa_125_vvll}. We have applied the same QT approach to $\PH\to\PZ(\nu_{\Pe}\bar{\nu}_{\Pe})\,{\mu^+}{\mu^-}$, as done for the case with four charged leptons, finding very similar results, \emph{i.e.}~a negative $\rho$ matrix.  
It is worth noting that this signature has an on-shell boson decaying to neutrinos, while the off-shell pair is still formed by charged leptons, therefore allowing for an intermediate photon at NLO, as in the four-charged-lepton case.

Following the same procedure for $\PH\to \PZ(\Pe^+\Pe^-)\nu_\mu \bar{\nu}_\mu$ the NLO spin-density matrix reads,
\begin{equation*}
\label{NLOrho10eevv}
\begin{pmatrix}
\cdot & \cdot& \cdot& \cdot& \cdot& \cdot& \cdot& \cdot& \cdot\\
 \cdot& \cdot & \cdot& \cdot & \cdot& \cdot& \cdot& \cdot& \cdot\\
 \cdot& \cdot& \blue{0.057} & \cdot& \blue{-0.225} & \cdot& \blue{0.057} & \cdot& \cdot\\
 \cdot& \cdot & \cdot& \cdot & \cdot& \cdot& \cdot& \cdot& \cdot\\
 \cdot& \cdot& \blue{-0.225} & \cdot& \blue{0.886} & \cdot& \blue{-0.225} & \cdot& \cdot\\
 \cdot& \cdot& \cdot& \cdot& \cdot& \cdot & \cdot& \cdot & \cdot\\
 \cdot& \cdot& \blue{0.057} & \cdot& \blue{-0.225} & \cdot& \blue{0.057} & \cdot& \cdot\\
 \cdot& \cdot& \cdot& \cdot& \cdot& \cdot & \cdot& \cdot & \cdot\\
 \cdot& \cdot& \cdot& \cdot& \cdot& \cdot& \cdot& \cdot& \cdot\\
\end{pmatrix} \,, \nonumber
\end{equation*}
which is positive semidefinite
within numerical accuracy and remains close
to its LO form.
The different outcomes of these two tests identify the
off-shell charged-lepton current as the source of the
reconstruction failure: changing the on-shell decay does
not restore positivity, whereas replacing the off-shell
charged-lepton pair by neutrinos removes the photon-mediated
current and restores a physical reconstruction.

The rank-one angular moments in the off-shell
neutrino channel still receive corrections from
the charged-lepton analyser on the on-shell side.
These are accounted for by the improved
$\eta_\ell(\MZ)$.
The neutrino current itself is purely chiral
in the massless limit, fixing its two-body
spin-analysing power to $\pm1$ (with sub-permille deviations from $\pm 1$ at NLO EW).

When the off-shell pair is charged, a loop-induced
$\PH\to\PZ\gamma^*$ contribution interferes with
the LO amplitude and becomes particularly
important at low virtualities.
The stability of the off-shell neutrino control
channel also suggests that loop topologies
without an $s$-channel $\PZ$ or photon coupling
to that current are numerically subleading.
This is indirect evidence, since the two channels
also differ in their remaining loop contributions.
The tree-level SMEFT example considered next
provides a complementary test in which the
relative strengths of the neutral-current
contributions can be varied.

\subsection{Tree-level modeling with the SMEFT}\label{sec:SMeft}
The SMEFT provides a tree-level test of the mechanism
identified in the preceding sections.
Effective Higgs--gauge interactions generate both
$\PH\to\PZ\PZ^*$ and $\PH\to\PZ\gamma^*$
contributions, whose relative importance can be
varied consistently through the Wilson coefficients.
This allows us to isolate the ingredients responsible
for the reconstruction failure and study its dependence
on the relative strengths of the two contributions.
Such freedom is absent in the SM NLO calculation,
where their relative weights are fixed once the
SM parameters and kinematics are specified.
The tree-level example therefore provides a controlled
setting in which to explore the mechanism beyond
the particular numerical configuration realised
by SM radiative corrections.

We consider
$\PH\to\PZ(\ell^+\ell^-)\ell'^+\ell'^-$
in the single-NWA description, retain SM leptonic
decay currents, and neglect contact interactions
involving leptons.
For this illustrative test, we restrict the
dimension-six SMEFT to the Warsaw-basis operators
\cite{Grzadkowski:2010es},
\beq
Q_{\varphi W}=\varphi^\dagger \varphi W_{\mu\nu}^iW^{\mu\nu,i},\,\quad
Q_{\varphi B}=\varphi^\dagger \varphi B_{\mu\nu}B^{\mu\nu}\,,
\eeq
with Wilson coefficients $C_{\varphi W}$ and
$C_{\varphi B}$, respectively.
Within these assumptions, the amplitude reads:
\begin{align}
\mc A \propto&
-\left(
\frac{g_{\mu\nu} }2+
\frac{ c_w^2 C_{ \varphi W}+s_w^2 C_{\varphi B}}{\Lambda^2{G_{\rm F}\MZ^2}}{\mc T}_{\mu\nu}
\right)
\frac{
J^\mu_{\PZ}(\MZ) J^\nu_{\PZ}(Q)
}{Q^2-\MZ^2}
\nonumber\\
&+
\sqrt{2}
\frac{s_w^2 c_w^2(C_{\varphi W} -C_{\varphi B})}{\Lambda^2{G_{\rm F}}\MZ^2}{\mc T}_{\mu\nu}
\frac{
J^\mu_{\PZ}(\MZ)J^\nu_{\gamma}(Q)
}{Q^2}
\label{eq:smeftmastereq}
\end{align}
where we have defined:
\begin{align}
   & c_w=\frac{\MW}{\MZ},\quad s_w = \sqrt{1-c_w^2}\,,\nonumber\\
   &
   J^\mu_{V}(Q_V)
   ={\overline{\psi}_i(\ell^-)\gamma^\mu\left(g^{(V)}_V-g^{(V)}_A\gamma_5\right)\psi_j(\ell^+)}\,,\nonumber\\
&    {\mc T}_{\mu\nu} = p^{(\PH)}_\mu p^{(\PH)}_\nu - g_{\mu\nu}\frac{\MH^2-\MZ^2-Q^2}2\,.\nonumber
\end{align}For both the off-shell $\PZ$ and the timelike
virtual photon, current conservation allows the
propagator numerator to be decomposed into three
spin-one polarisation components,
$\lambda=0,\pm1$.
The term proportional to the dilepton momentum
vanishes upon contraction with the massless
leptonic current.
Thus both contributions admit a qutrit helicity
description individually, and
Eq.~\eqref{eq:smeftmastereq} takes the form
\begin{align}
 \mc A &=\sum_\lambda\left[\mc A_{\lambda}^{(\PZ)}\left(\varepsilon^{*\,\alpha}_{\lambda}J_{\PZ,\alpha}\right) +\mc A_{\lambda}^{(\gamma)}\left(\varepsilon^{*\,\alpha}_{\lambda}J_{\gamma,\alpha}\right)\right]\nonumber\\
&=\sum_\lambda\left[\mc A_{\lambda}^{(\PZ)}\mc D^{(\PZ)}_{\lambda} +\mc A_{\lambda}^{(\gamma)}\mc D^{(\gamma)}_{\lambda}\right]\,,\label{eq:smeftampbis}
\end{align}
where we have defined $\mc D^{(V)}_{\lambda} = \varepsilon^{*\,\alpha}_{\lambda}J_{V,\alpha}$.
For a single intermediate current, the familiar
production--decay representation follows from
the factorisation of the amplitude into
$\mc A_\lambda$ and $\mc D_\lambda$.
After summing over unobserved degrees of freedom
and integrating over the on-shell $\PZ$ decay
angles, the corresponding single-system
density matrix is defined through
\begin{equation}\label{eq:rhoVdef}
    \frac{\rd \sigma}{\rd \Omega} \propto |\mc A|^2=\sum_{\lambda,\lambda'}\mc A_\lambda \mc A^*_{\lambda'}\mc D_\lambda \mc D^*_{\lambda'}\propto
    \sum_{\lambda,\lambda'}\rho_{\lambda\lambda'}\Gamma_{\lambda\lambda'}\,.
\end{equation}
From Eq.~\eqref{eq:smeftampbis} we get,
\begin{eqnarray}
    \frac{\rd \sigma}{\rd \Omega} 
    &\propto&\sum_{\lambda,\lambda'}
     \underbrace{\mc A_{\lambda}^{(\PZ)}\mc A_{\lambda'}^{*(\PZ)}
     }_{\propto\, \rho^{(\PZ)}_{\lambda\lambda'}}
    \underbrace{\mc D^{(\PZ)}_{\lambda}
    \mc D^{*(\PZ)}_{\lambda'}}_{\propto\, \Gamma^{(\PZ)}_{\lambda\lambda'}} \nonumber\\
&+&    2 \,{\rm Re}
    \sum_{\lambda,\lambda'}
    \mc A_{\lambda}^{(\PZ)}\mc A_{\lambda'}^{*(\gamma)}\mc D^{(\PZ)}_{\lambda} 
    \mc D^{*(\gamma)}_{\lambda'}\nonumber\\
    &+&
     \sum_{\lambda,\lambda'}
     \underbrace{\mc A_{\lambda}^{(\gamma)}\mc A_{\lambda'}^{*(\gamma)}
     }_{\propto\, \rho^{(\gamma)}_{\lambda\lambda'}}
     \underbrace{\mc D^{(\gamma)}_{\lambda}
    \mc D^{*(\gamma)}_{\lambda'}}_{\propto\, \Gamma^{(\gamma)}_{\lambda\lambda'}}\label{eq:3liner}
    \,,    
\end{eqnarray}
The first and third terms separately admit the
production--decay representation of
Eq.~\eqref{eq:rhoVdef}, but involve different decay
matrices.
The interference term instead contains a
transition decay matrix built from
$\mc D^{(\PZ)}_\lambda
\mc D^{*(\gamma)}_{\lambda'}$.
The full distribution therefore involves several
channel-dependent production and decay structures.

The obstruction to standard tomography is the
simultaneous presence of different production
tensors and different chiral decay currents.
In particular, the $\PZ$ contribution contains
both the SM and dimension-six production tensors,
whereas the photon contribution contains only
the latter.
At the same time, $\PZ^*$ and $\gamma^*$ couple
differently to the lepton helicities.
Their sum consequently does not, in general,
factorise into a single production density matrix
and a universal decay analyser.

This issue is not restricted to coherent
interference.
Even an incoherent sum of channels with different
production states and decay analysers need not
admit a single universal analyser.
The angular coefficients remain measurable in either
case,
but their
conversion into entries of the spin-density matrix is not guaranteed to be universal, 
\emph{i.e.} the existence of a unique mapping between measured angular coefficients and the spin-density-matrix entries is not given.
If, however, we had an incoherent sum of channels with different production states but the same decay analyser, or a single channel with two different decay analysers, then the conversion would work. It is the combination of these two effects that causes the problem.

Section~\ref{sec:decaymatrix} identifies the
rank-two sector for which that conversion
nevertheless remains universal, and also provides a 
generalised strategy to fit interference terms in the structure 
of Eq.~(\ref{eq:rhoVdef}).

An experimental description could instead use
templates that retain the channel-dependent
angular structures and their interference.
This is similar in spirit to polarisation-template
analyses
\cite{Aaboud:2019gxl,CMS:2020etf,ATLAS:2022oge,
ATLAS:2023zrv,ATLAS:2024qbd,ATLAS:2025wuw,
ATLAS:2026kzu}, but requires 
to assume the SM amplitude in the interference term, therefore fixing the $\rho$ entries.

We illustrate this mechanism at $Q=10\GeV$
with $C_{\varphi W}/\Lambda^2=-1\TeV^{-2}$
and $C_{\varphi B}/\Lambda^2=0$.
Applying the reconstruction prescription used
above to the tree-level prediction, initially
retaining only terms linear in the Wilson
coefficients, gives the following operator:
\begin{equation*}
\label{SMEFTrho}
\begin{pmatrix}
\teal{~-0.030} & \cdot& \cdot& \cdot& \cdot& \cdot& \cdot& \cdot& \cdot\\
 \cdot& \cdot & \cdot& \cdot & \cdot& \cdot& \cdot& \cdot& \cdot\\
 \cdot& \cdot& \blue{0.073} & \cdot& \blue{-0.202} & \cdot& \blue{0.044} & \cdot& \cdot\\
 \cdot& \cdot & \cdot& \cdot & \cdot& \cdot& \cdot& \cdot& \cdot\\
 \cdot& \cdot& \blue{-0.202} & \cdot& \blue{0.913} & \cdot& \blue{-0.202} & \cdot& \cdot\\
 \cdot& \cdot& \cdot& \cdot& \cdot& \cdot & \cdot& \cdot & \cdot\\
 \cdot& \cdot& \blue{0.044} & \cdot& \blue{-0.202} & \cdot& \blue{0.073} & \cdot& \cdot\\
 \cdot& \cdot& \cdot& \cdot& \cdot& \cdot & \cdot& \cdot & \cdot\\
 \cdot& \cdot& \cdot& \cdot& \cdot& \cdot& \cdot& \cdot& \teal{-0.030~} \\
\end{pmatrix}\,, \nonumber
\end{equation*}
The negative diagonal entries establish that
this operator is not positive semidefinite.
The failure persists when squared dimension-six
amplitudes are included.
This is an essential diagnostic: the square of
the retained tree-level amplitude defines a
non-negative angular distribution, yet its
inversion through the assumed analyser still
produces an unphysical operator.
The inclusion of these quadratic terms is used
here to test the reconstruction map, rather
than to provide a complete SMEFT prediction
through order $\Lambda^{-4}$.

As a control, we set
$C_{\varphi B}/\Lambda^2
=C_{\varphi W}/\Lambda^2=-1\TeV^{-2}$.
The $\PH\PZ\gamma$ contribution in
Eq.~\eqref{eq:smeftmastereq} then vanishes.
Within the single-NWA process considered here,
the deformation affects only the
$\PH\PZ\PZ$ production vertex, leaving a single
leptonic decay analyser.
The reconstructed matrix is correspondingly
close to its SM form:
\begin{equation*}
\label{SMEFTrhoHZZ}
\begin{pmatrix}
\cdot & \cdot& \cdot& \cdot& \cdot& \cdot& \cdot& \cdot& \cdot\\
 \cdot& \cdot & \cdot& \cdot & \cdot& \cdot& \cdot& \cdot& \cdot\\
 \cdot& \cdot& \blue{0.060} & \cdot& \blue{-0.230} & \cdot& \blue{0.060} & \cdot& \cdot\\
 \cdot& \cdot & \cdot& \cdot & \cdot& \cdot& \cdot& \cdot& \cdot\\
 \cdot& \cdot& \blue{-0.230} & \cdot& \blue{0.880} & \cdot& \blue{-0.230} & \cdot& \cdot\\
 \cdot& \cdot& \cdot& \cdot& \cdot& \cdot & \cdot& \cdot & \cdot\\
 \cdot& \cdot& \blue{0.060} & \cdot& \blue{-0.230} & \cdot& \blue{0.060} & \cdot& \cdot\\
 \cdot& \cdot& \cdot& \cdot& \cdot& \cdot & \cdot& \cdot & \cdot\\
 \cdot& \cdot& \cdot& \cdot& \cdot& \cdot& \cdot& \cdot& \cdot \\
\end{pmatrix}\,, \nonumber
\end{equation*}
Once again, this holds truncating the SMEFT expansion both at linear and at quadratic level.

The comparison isolates the role of the
photon-mediated current.
A modified $\PH\PZ\PZ$ production tensor alone
does not invalidate the universal decay map,
whereas the additional $\PH\PZ\gamma^*$
contribution can do so through its different
production and decay structures.
The persistence of the failure after including
squared dimension-six amplitudes demonstrates
that it is not solely a truncation artefact.

This tree-level example therefore supports
the interpretation of the SM NLO results
in terms of the coexistence of
$\PZ^*$- and $\gamma^*$-mediated currents.
In the full off-shell process, additional
$\PH\gamma\gamma$ contributions, generated
at tree level in the SMEFT and at one loop
in the SM, introduce further channel-dependent
structures.

\subsection{Generalised transition decay matrices}\label{sec:decaymatrix}
As anticipated in Sect.~\ref{sec:SMeft}, the conversion of angular coefficients into spin-density-matrix entries is not necessarily universal. We show that this comes uniquely from the 
rank-one sector of the decay matrix.
The universality of the rank-two sector follows from
a non-trivial property of the transition decay matrices.
For a neutral boson $a$ ($a=\gamma,\PZ$) that couples to 
massless-lepton currents with left(right)-chiral coupling strengths $c^{a}_{L(R)}$, the decay amplitude reads:
\begin{align}
\mc D_{\lambda, h}^{(a)}(\Omega) = \varepsilon_{\lambda}^{\mu,*}\,\bar{\psi}({\ell^-})\gamma_\mu \left(c^{a}_L\frac{1-\gamma_5}2 +c^{a}_R\frac{1+\gamma_5}2\right)\psi({\ell^+})\,.\nonumber
\end{align}
In order to recast the second line of Eq.~(\ref{eq:3liner})
into the structure of Eq.~(\ref{eq:rhoVdef}), namely,  $\sum_{\lambda,\lambda'}\rho^{ab}_{\lambda\lambda'}\,\Gamma^{ab}_{\lambda\lambda'}\,,$
we define a generalised version of transition decay matrices $\Gamma_{\lambda\lambda'}$,
\begin{align}
\Gamma^{ab}_{\lambda\lambda'}(\Omega)
=
\frac{\displaystyle\sum_{h=L,R}\mc 
D^{(a)}_{\lambda, h}(\Omega) \mc D^{(b)*}_{\lambda',h}(\Omega)}{\displaystyle
\sum_{\lambda=0,\pm}\sum_{h=L,R}\mc 
D^{(a)}_{\lambda, h}(\Omega) \mc D^{(b)*}_{\lambda,h}(\Omega)
}\,,
\end{align}
which has rank-zero and rank-two components proportional
to the same chiral sum $S_{ab}$,
\begin{align}
S_{ab}=c_L^a c_L^{b*}+c_R^a c_R^{b*}\,,
\end{align}
whereas its rank-one
component is proportional to $D_{ab}$, 
\begin{align}
D_{ab}=c_L^a c_L^{b*}-c_R^a c_R^{b*}\,.
\end{align}
Equivalently, its angular decomposition takes the form
\begin{equation}
\Gamma^{ab}(\Omega)
=
{\cal K}_0(\Omega)+{\cal K}_2(\Omega)
+\frac{D_{ab}}{S_{ab}}{\cal K}_1(\Omega),
\end{equation}
where ${\cal K}_l$ are channel-independent geometrical
kernels of rank-$l$.
The essential point is that the rank-zero and rank-two
terms carry the same coupling factor.
Consequently, normalising to the scalar component
also fixes the rank-two analyser, independently of
whether the contribution is $\PZ\PZ$, $\gamma\gamma$,
or a $\PZ\gamma$ interference term.
The rank-one analyser instead depends on the
channel-dependent ratio $D_{ab}/S_{ab}$ whenever
$S_{ab}\neq0$.
The unnormalised relation above remains valid also
when $S_{ab}=0$.
Thus the coherent sum over neutral currents preserves
the universal rank-two conversion factor ($\xi_2^{(i)}$ in Eq.~(\ref{eq:gamma_to_C})) between angular coefficients and spin-density-matrix entries,
\begin{equation}\label{eq:leq2rel}
{\sqrt{40\pi}}\,\alpha^{(i)}_{2m}= A^{(i)}_{2m},
\qquad
40\pi\,\gamma_{2m2m'}=  C_{2m2m'}\,,
\end{equation}
without requiring a universal rank-one spin-analysing power.
This does not imply that the
rank-two moments themselves receive no radiative
corrections.

\subsection{Entanglement markers}\label{sec:Cbound}

The limitations of the full spin-density-matrix
reconstruction do not preclude the construction of
entanglement markers from partial angular information. 
However, we will show in the following that what we have discussed so far has non-trivial implication for this purpose. In particular, only under certain assumptions it is possible to employ the same entanglement markers that are valid at LO.
The recent ATLAS analysis \cite{ATLAS:2026hye}
reported evidence for spin entanglement in $\PZ\PZ^*$ pairs from
Higgs-boson decays, and measured the angular coefficients $C_{212-1}$ and $C_{222-2}$. According to the Peres-Horodecki criterion \cite{Peres:1996dw,Horodecki:1996nc}, the condition
\beq\label{eq:ATLAScond}
C_{222-2} \neq 0\quad \textrm{ or } \quad C_{212-1}\neq 0\,,
\eeq
is sufficient and necessary to have an entangled state in the decay of a scalar to two qutrits \cite{Aguilar-Saavedra:2022wam}, however, its validity and interpretation relies on assumptions about the structure of the underlying spin state.
Namely, it is mandatory that the spin-density matrix has the same form of the LO one, which is possible only if  the LO relations involving the $A$ and $C$ coefficients are valid.  In order to verify such conditions, $C_{111-1}$ and $C_{1010}$ have to be extracted, but as discussed at lengths in the previous sections, it is not possible for the $\PH\to \PZ(\Pe^+\Pe^-)\,\mu^+\mu^-$ decay.

The only possible alternative is therefore to use an entanglement
criterion that applies to mixed two-qutrit states
and requires only the reliably accessible coefficients: rank-zero and rank-two spin-density coefficients.
Indeed, Section~\ref{sec:decaymatrix} establishes that the rank-zero and rank-two components of the transition
decay matrices carry the same coupling factor.
Their common normalisation preserves the universal
conversion between rank-two angular moments and
spin-density coefficients, even when no universal
rank-one analyser exists.

This property allows us to construct a conservative
concurrence-based bound using only rank-two
correlations, as already shown in \citere{DelGratta:2025qyp}.
The omitted rank-one correlation coefficients enter the
underlying purity bound through non-negative
squared moduli. Dropping these terms therefore
weakens the bound without requiring them to vanish
or to be reconstructed:
\begin{align}\label{eq:CLB2}
\mc C_{LB}^{l>1}=\frac29\bigg\{&-2 + \displaystyle
\sum_{m=-2}^2
\sum_{m'=-2}^2\Big|C_{2m2m'}\Big|^2 \\
&+{\rm max}\bigg[
\displaystyle\sum_{l=1}^2\sum_{m=-l}^{+l}
\left(-2\left|{A^{(1)}_{lm}}\right|^2
+\left|{A^{(2)}_{lm}}\right|^2\right),
\nonumber\\
&
\hspace{1.3cm}\displaystyle\sum_{l=1}^2
\sum_{m=-l}^{+l}
\left(\left|{A^{(1)}_{lm}}\right|^2
-2\left|{A^{(2)}_{lm}}\right|^2\right)
\bigg]
\bigg\}\,.\nonumber
\end{align}
With the standard definition of concurrence,
this expression provides a lower bound on its square.
A positive value is sufficient to certify entanglement,
whereas a non-positive value is inconclusive
\cite{Zhang:2008ebu}. In other words, a positive value of $\mc C_{LB}^{l>1}$  is a sufficient but not necessary condition for entanglement.


The definition in Eq.~\eqref{eq:CLB2} was provided in Ref.~\cite{DelGratta:2025qyp} for avoiding rank-one coefficients in the case of the Higgs decay to four leptons. However,  such definition is valid for a general two-qutrit system, without assuming the relations that are specific for the Higgs decay to four leptons. If further assumptions are made, equation~\eqref{eq:CLB2} can be  simplified. 
In the case of Higgs decay to four leptons
the $\alpha^{(i)}_{1m}, m=0,\pm1$ and $\alpha^{(i)}_{2m}, m\neq0$  coefficients vanish by symmetry \cite{DelGratta:2025qyp}, like the $\gamma_{2m2m'}$ coefficients with $m\ne -m'$. Thus, Eq.~\eqref{eq:CLB2} 
can be expressed in terms of only the non-vanishing angular coefficients $\alpha^{(i)}_{20}$ and $\gamma_{2m2-m}$, which are directly measurable observables at colliders:
\begin{align}\label{eq:CLB2_alpha}
\mc C_{LB}^{l>1} = \frac29\bigg\{&-2 + (40\pi)^2\displaystyle
\sum_{m=-2}^2
\Big|\gamma_{2m2-m}\Big|^2 \\
&+(40\pi)\,{\rm max}\bigg[
\displaystyle
\left(-2\left|{\alpha^{(1)}_{20}}\right|^2
+\left|{\alpha^{(2)}_{20}}\right|^2\right),
\nonumber\\
&
\hspace{2.15cm}\displaystyle
\left(\left|{\alpha^{(1)}_{20}}\right|^2
-2 \left|{\alpha^{(2)}_{20}}\right|^2\right)
\bigg]
\bigg\}\,.\nonumber
\end{align}
Since Eqs. \eqref{eq:relA} and \eqref{eq:relC} can be recast in terms of angular coefficients  as
\begin{eqnarray}
\alpha^{(1)}_{20} = \alpha^{(2)}_{20} = \alpha_{20} \ ,  \nonumber \\
\gamma_{2020} = \frac{1}{40 \pi} - \frac{1}{\sqrt{80 \pi}} \alpha_{20} \ , \nonumber \\
\gamma_{222-2} = \frac{1}{40 \pi} + \frac{1}{\sqrt{80 \pi}} \alpha_{20}\,, \label{eq:alfagammaconditions}
\end{eqnarray}
if such conditions are satisfied, the definition of $C_{LB}^{l>1}$ can be further simplified as:
\begin{align}
\mc C_{LB}^{l>1} &= \frac{2 (40\pi)^2}{9} (|\gamma_{222-2}|^2 + 2 |\gamma_{212-1}|^2) \nonumber \\
 &= \frac{2 }{9} (|C_{222-2}|^2 + 2 |C_{212-1}|^2) \ . \label{eq:finalCLB}
\end{align} 

As can be seen, requiring $\mc C_{LB}^{l>1} $ as defined in Eq.~\eqref{eq:finalCLB} is equivalent to Eq.~\eqref{eq:ATLAScond}. However, while in the case of the LO picture Eq.~\eqref{eq:ATLAScond} is a {\it sufficient and necessary} condition for entanglement, when going to higher orders it is only a {\it sufficient} condition and only if Eqs.~\eqref{eq:alfagammaconditions} are all valid.

As a concrete example, the angular coefficients
in the upper block of
Tab.~\ref{tab:qtom_nwa_125_vvll}, obtained for
$\PH\rightarrow\PZ(\Pe^+\Pe^-)\mu^+\mu^-$
in the single NWA at $Q=10\GeV$, give
\begin{align}
\mc C_{LB}^{l>1}(\rm LO) =0.208(2)\,,\quad
\mc C_{LB}^{l>1}(\rm NLO_{\rm virt}) =0.195(2)\,,
\nonumber
\end{align}
showing that the marker remains positive and
inherits the perturbative stability of the
rank-two sector.

We stress that requiring $\mc C_{LB}^{l>1} > 0$ as proof of entanglement avoids completely the necessity of reconstructing the spin-density matrix and therefore extracting the problematic rank-one tensors. Using the definition in Eq.~\eqref{eq:CLB2_alpha}, is sufficient to measure the angular coefficients $\alpha$ and $\gamma$, while for the definition in Eq.\eqref{eq:finalCLB}, which is equivalent to Eq.~\eqref{eq:ATLAScond}, is necessary to verify the validity of the Eqs.~\eqref{eq:alfagammaconditions}.

\section{Conclusions}\label{sec:conclusion}
We have examined the limits of standard two-qutrit quantum
tomography in Higgs-boson decays to four charged leptons
beyond leading order.
Exact off-shell predictions, together with calculations
in the single-$\PZ$ narrow-width approximation, show that
the reconstruction of NLO EW angular distributions can
yield a Hermitian, unit-trace spin-density matrix that is not positive
semidefinite, even with perturbatively improved spin-analysing
powers.
The angular moments themselves remain well defined;
what fails is their interpretation through the assumed
production density matrix and universal decay analysers.

The reconstruction problem persists when higher-rank
angular contributions are negligible, resolved-photon
effects are suppressed, and the off-shell dilepton
virtuality $Q$ is fixed.
It therefore cannot be attributed solely to additional
angular structures, real radiation, or integration over
the physical $Q$ range.
The inclusion of squared one-loop contributions provides
a further test of the role of perturbative truncation.

We trace the central obstruction to the coexistence of
$\PZ^*$- and $\gamma^*$-mediated contributions to the
off-shell dilepton current.
Their different couplings to the Higgs boson and to
leptons lead to distinct production and decay structures,
whose coherent combination does not, in general, admit
the universal factorisation assumed in standard tomography.
Photon-mediated effects are particularly pronounced
at small dilepton virtualities.
Replacing the off-shell charged-lepton pair by neutrinos
supports this interpretation: in
$\PH\to\PZ(\ell^+\ell^-)\nu\bar{\nu}$, the photon-mediated
off-shell current is absent and the reconstructed matrix
remains positive semidefinite and perturbatively stable.
The stability of this control channel also suggests
that loop topologies without an $s$-channel $\PZ$
or photon coupling to the off-shell current are
numerically subleading in the observables considered.
Although such contributions can introduce additional
non-factorising structures, this provides indirect
evidence that the coexistence of $\PZ^*$- and
$\gamma^*$-mediated amplitudes is the dominant source
of the reconstruction failure in the four-charged-lepton
channel.

The tree-level SMEFT example provides a simple,
tunable setting in which to test this interpretation.
By varying the Wilson coefficients, the relative
strengths of the $\PZ^*$- and $\gamma^*$-mediated
contributions can be adjusted consistently,
allowing the reconstruction mechanism to be explored
beyond the particular balance realised in the
SM NLO calculation, where these strengths are fixed
once the SM parameters and kinematics are specified.
An effective $\PH\PZ\gamma$ interaction reproduces
the reconstruction failure even when squared
dimension-six amplitudes are included, whereas
 a $\PH\PZ\PZ$-only deformation 
is not problematic.
This comparison demonstrates that the obstruction
is not solely a perturbative-truncation artefact
and provides a controlled framework for testing
a wider range of
neutral-current contributions.

This picture is consistent with results in related
processes.
The same mechanism is absent in the charged-current
decay $\PH\to\ell^+\nu_\ell\ell'^-\bar{\nu}_{\ell'}$
\cite{DelGratta:2025qyp,Goncalves:2025xer}, which has
no analogous photon-mediated dilepton current.
In $\Pp\Pp\to\PZ\PZ$, the dominance of the doubly
resonant region suppresses photon-mediated and
non-resonant contributions, and much of the EW correction
can be absorbed into the spin-analysing powers
\cite{Grossi:2024jae,Goncalves:2025mvl,DelGratta:2025xjp}.
Related limitations also arise in
$\PH\to\ell^+\ell^-\Pj\Pj$, despite generally smaller
NLO corrections \cite{Goncalves:2026njf}.
The scope of our analysis extends beyond the
four-charged-lepton final state.
The framework developed here, the interpretation
in terms of channel-dependent production and decay
structures, and the proposed reconstruction strategy
apply more generally to Higgs-boson decays to two
fermion pairs, with the appropriate decay analysers
and under the assumptions stated above.
Other channels share only some of the features
responsible for the particularly pronounced
reconstruction failure in $4\ell$.
The size of the effects and the validity of
a full reconstruction therefore depend on the channel,
while the procedure to test the goodness of the quantum tomography
is general.

The Higgs-boson decay to four charged leptons, despite an appealingly clean signature at colliders, is particularly problematic due to a combination of various effects, namely the off-shell nature of one $\PZ$ boson, the presence of a photon-mediated contribution, and the almost vanishing value of the spin-analysing power due to accidental cancellations.
Nonetheless, we show that partial information of the spin-density matrix is accessible despite the failure of
the full reconstruction.
The entanglement conditions $C_{222-2}\neq 0$ or $ C_{212-1}\neq 0$ \cite{Aguilar-Saavedra:2022wam} are necessary and sufficient only under strong assumptions on the spin-density matrix. 
Such assumptions are indeed satisfied at tree level in the SM. In general, however, they can be checked only via the extraction of the experimentally inaccessible rank-one coefficients
Nonetheless, it is possible to construct a measurable lower bound on the squared concurrence of two-qutrit states that does not depend on such inaccessible rank-one coefficients and, if positive, implies entanglement. In addition, if some relations involving the measurable rank-two angular distributions are satisfied (see Eq.~\eqref{eq:alfagammaconditions}),  the positivity of such lower bound is equivalent to the condition $C_{222-2}\neq 0$ or $ C_{212-1}\neq 0$.  
In this case such conditions are sufficient but not necessary. 


In summary, these results set clear limits on the interpretation
of reconstructed quantum observables.
A more general tomography must account for
channel-dependent production and decay structures,
potentially by retaining additional unresolved degrees
of freedom.
Fits incorporating these structures offer one possible
route, while alternative descriptions based on fermion
spins or orbital angular momentum merit further study.
The process thus remains valuable to investigate both quantum-entanglement and sensitivity to new-physics effects.

\begin{acknowledgments}
We are grateful to Federica Fabbri, Alessandro Vicini and Eleni Vryonidou for  useful comments on the manuscript.
The authors acknowledge support from the COMETA EU COST Action (CA22130).
MDG and DP acknowledge the financial support by the MUR (Italy), with funds of the European Union (NextGenerationEU), through the PRIN2022 grant 2022EZ3S3F.
GP is funded by the EU Horizon Europe research and innovation programme under the Marie-Sk\l{}odowska Curie Action (MSCA) ``POEBLITA - POlarised Electroweak Bosons at the LHC with Improved Theoretical Accuracy'', grant agreement Nr.~101149251 (CUP H45E2300129000). 
\end{acknowledgments}

\bibliographystyle{JHEP}
\bibliography{hqtom}

\end{document}